\documentclass[showpacs,preprint]{revtex4}
\usepackage{mathrsfs}
\usepackage{amssymb}
\usepackage{latexsym}
\usepackage{hyperref}
\usepackage{amsmath}

\begin{document}

\title{Thermodynamics for Kodama observer in general spherically symmetric
spacetimes}
\author{Yi-Xin Chen$^{1}$}\email{yxchen@zimp.zju.edu.cn}
\author{Jian-Long Li$^{1}$}\email{marryrene@gmail.com}
\author{Yong-Qiang Wang$^{2}$}\email{yqwang@lzu.edu.cn}

\affiliation{$^{1}$Zhejiang Institute of Modern Physics, Zhejiang
University, Hangzhou 310027, China}
\affiliation{$^{2}$Institute of Theoretical Physics, Lanzhou University
Lanzhou 730000, China}

\begin{abstract}
By following the spirit of arXiv:1003.5665, we define a new Tolman temperature of Kodama observer directly related to its acceleration. We give a generalized integral form of thermodynamics relation on virtual sphere of constant $r$ in non-static spherical symmetric spacetimes. This relation contains work term contributed by `redshift work density', `pressure density' and `gravitational work density'. We illustrate it in RN black hole, Dilaton-Maxwell-Einstein black hole and Vaidya black hole. We argue that the co-moving observers are not physically related to Kodama observers in FRW universe unless in the vacuum case. We also find that a generalized differential form of first law is difficult to be well defined, and it would not give more information than the integral form.
\end{abstract}
\maketitle

\section{Introduction}
Black hole thermodynamics is one of the most interesting discovery in general relativity, which first shows an implication of the analogue
relation between gravity and thermodynamics \cite{Bekenstein,Bardeen}.
Recently, a radical idea is stimulated that gravity might be originated from
the thermodynamics of the unknown microstructure of spacetime.
Jacobson \cite{Jacobson, Eling} first illustrates it by obtaining Einstein's
equation from the first law of thermodynamics, $\delta Q= T dS$,
which are defined on a local Rindler horizon. This illustration can
be extended to non-Einstein gravity(for a review, see \cite{10}). In \cite{Pad3, Pad1, Pad2, Pad4, Pad5, Pad6}, Padmanabhan asserts that `spacetime can be heat up', and the thermodynamics exists in static background of various gravity theories off the black hole horizon as an equipartition rule. In contrary, another illustration is the entropic force proposed by Verlinde \cite{Verlinde}. In his work, spacetime is the place where information is stored, and the fundamental unit of 3-dimensional space is 2-dimensional holographic screen. Defining thermodynamical properties on the holographic screens by using Killing observer, and applying the principle of holography, Einstein equation are derived from the thermodynamics in static case. It can be regarded as a kind of reverse logic to \cite{Pad3}. Namely, the origin of the gravity is the entropic force. An incomplete list of thermodynamics topics in entropic force frame is \cite{Zhao, Banerjee:2010yd, Kiselev, Pan, Nicolini, Fursaev1, Piazza, Chen:2010ay, Tian, Banerjee:2010ye}.

The progresses \cite{Pad3, Verlinde} rely on the thermodynamical variables defined by Killing observer, which is absent in time-dependent case. Since our universe is globally a non-static solution of gravity equation, the thermodynamics analogy of gravity in time-dependent spacetime is more profound. This difficulty might be alleviated a little by a geometrically natural divergence-free preferred vector field proposed by Kodama \cite{Kodama} in 1980. The so-called ``Kodama vector'' defines a natural timelike direction in general spherical symmetrical spacetime, and induces an unexpected conserved charge, which coincides with Misner-Sharp energy\cite{Misner}. Over the years, several discussions of thermodynamics by Kodama observers associated to Kodama vector are made. For example, Hayward \emph{et al} \cite{Hayward0, Hayward1, Hayward2, Hayward3} define the Hawking temperature \cite{Hawking, Unruh} of dynamical black holes and the generalized first law of thermodynamics. This Hawking temperature are supported by tunneling method \cite{Wilczek}, in which the Killing observer is replaced by Kodama observer \cite{CriscienzoPLB, CriscienzoCQG, Cai:2008gw, Chen:2010hc}. Kodama observer is also helpful for generalization of thermodynamics in FRW universe \cite{Cao, Cai:2008gw, Chen:2010hc, Cai:2010hk, Cao3, Akbar, Zhu, Wu10}, dynamical problems \cite{Cao1, Abreu:2010ru, Csi, Racz} and entropy bound \cite{Abreu:2010sc}.

As Kodama observer is a good substitute of Killing observer in non-static background, it is applied to extend \cite{Verlinde} into dynamical case. In \cite{Cao1}, applying Hawking temperature and the generalized first law defined by Hayward, Cai \emph{et al} give the formulism of entropic force in general spherical symmetric spacetimes. Most of the formulism are defined on the trapping horizon. They also define a new Newton potential by Kodama vector. By following the same spirit of Hayward and Verlinde, Wu \cite{Wu:2010ty} \emph{et al} obtain a new temperature and first law on holographic screens. However, the thermodynamics above are mainly in a differential form, while the works \cite{Pad3, Verlinde} are mainly in an integral form. In order to investigate whether Einstein equation can be emergent from entropic force in dynamical background, we should first recast the gravity into thermodynamics in an integral form.

In this article, we extend the work of Padmanabhan \cite{Pad3} into general spherical symmetric spacetimes. Firstly, we choose the Kodama preferred time flow as the time coordinate, and redefine a diagonal metric without loss of generality. Secondly, we define a Kodama temperature as $T_K=N |a|/2\pi$, where $|a|$ is the magnitude of acceleration of Kodama observer, and N is the normalized factor. This temperature is different from Hayward's temperature and Wu's temperature, while the latter two are not directly related to the observer acceleration unless on the trapping horizon. Third, since Misner-Sharp mass are the conserved charge of Kodama vector, by using it we extend the thermodynamics based on Komar mass in \cite{Pad3} into general static case and non-static case. It can be viewed as the preparation of studying entropic force in non-static case. Our integral form of thermodynamics are numerically equivalent to the Smarr-like law in \cite{Cao1} on trapping horizon, but the physical interpretation of `work' terms in our work is more detailed and clear. Then we illustrate our method on RN black hole, Dilaton-Maxwell-Einstein black hole and Vaidya black hole. We also find that there is lack of physical relation between the co-moving observer and Kodama observer in FRW universe, unless it is a vacuum solution. In the end of the article, we discuss the consistency between the integral and differential form of thermodynamics.

The layout of the article is as follows: In Sec. 2 we present a quick review of Kodama vector and the preferred time flow. Next, in Sec. 3, we recast the Einstein gravity into a non-equipartition relation in static case, and we illustrate it in RN black hole, Dilaton-Maxwell-Einstein black hole. In Sec. 4, we extend it into non-static case, and give an example in Vaidya black hole. We also discuss the lack of physical relation between the co-moving observer and Kodama observer in FRW universe. In Sec. 5, we discuss the consistency between our conclusion and the earlier works. Lastly, we add a brief conclusion.

\section{Kodama observer and the preferred time flow}
It is known that there exist a special fiducial observer called Kodama observer in any time-dependent spherically symmetric metric,
\begin{equation}
ds^2=\gamma_{ij} dx^i dx^j+r^2(x) {d\theta^2 +\sin^2\theta d\phi^2}.\label{1}
\end{equation}
Here $x^i$ run over the radial-temporal plane, while $\theta$ and $\phi$ run over the surface of spherical symmetry. The discussion can also be generalized to $(d+1)$ dimension with $(d-1)$ dimensional spherical symmetry, or higher-dimensional generalizations in Gauss-Bonnet gravity, or Horava gravity \cite{Cai:2009ph}. But in our context, we focus on this $(3+1)$ dimensional case. The Kodama vector are given by,
\begin{equation}
K^i(x)=\frac{1}{\sqrt{-\gamma}} \epsilon^{ij} \partial_j r, \ K^\theta=0=K^\phi.
\end{equation}
Here $\epsilon^{ij}$ is the Levi-Civita tensor in the radial-temporal plane. Kodama observers are along these time-like curves with constant $r(x^i)$, which select a preferred time flow orthogonal to $r$ direction.
By using the Clebsch decomposition theorem \cite{Abreu:2010ru}, without loss of generality, we can always get a preferred coordinates with diagonal metric,
\begin{equation}
ds^2=-e^{-2\Phi(r,t)}N^2(r,t)dt^2 +\frac{1}{N^2(r,t)} dr^2 +r^2 d\Omega^2.\label{2}
\end{equation}
It can be regarded as spacetime re-foliated by Kodama preferred time flow. Then the Kodama vector is,
\begin{equation}
\begin{split}
K^\mu&=e^{\Phi(r,t)}(1, 0; 0, 0),\\
K_\mu&=-e^{-\Phi(r,t)}N^2(r,t)(1, 0; 0, 0),\\
K^\mu K_\mu&=-N^2(r,t).
\end{split}
\end{equation}

The normalized Kodama vector are given by,
\begin{equation}
\begin{split}
k^\mu&=\frac{K^\mu}{|K|}=e^{\Phi(r,t)}(\frac{1}{N(r,t)}, 0, 0, 0),\\
k_\mu&=\frac{g_{\mu\nu}K^\mu}{|K|}=-e^{-\Phi(r,t)}N(r,t)(1, 0, 0, 0).
\end{split}
\end{equation}
Notice that it also parallel to the unit normal vector $n^\mu$ of $t=$const hypersurface $\Sigma$,
\begin{equation}
k^\mu=n^\mu.
\end{equation}
The corresponding Kodama surface gravity is
\begin{equation}
\kappa_h=\frac{1}{2\sqrt{-\gamma}}\partial_i(\sqrt{-\gamma}\gamma^{ij}\partial_j r)|_h.
\end{equation}
Here $h$ denotes the trapping horizon. There are two conserved charged associated to Kodama observer,
\begin{equation}
\begin{split}
Q_1&=\int_\Sigma d^3 x \sqrt{h} g_{\mu\nu}K^\mu n^\nu=\int_0^{r=r_\Sigma} r^2 \sin\theta dr d\theta d\phi=\frac{4\pi}{3}r^3|_{r=r_\Sigma}=V,\\
Q_2&=\int_\Sigma d^3 x \sqrt{h} T_{\mu\nu}K^\mu n^\nu=\int_0^{r=r_\Sigma} r^2 \sin\theta T_{00} dr d\theta d\phi=E_{MS}|_{r=r_\Sigma}.
\end{split}
\end{equation}
Here $dx$ and $h_{ij}$ are the induced coordinate and metric on the $\Sigma$, $V$ is the coordinate volume and $E_{MS}$ is the Misner-Sharp energy,
\begin{equation}
E_{MS}=\frac{r}{2}(1-\gamma^{ij}\partial_i r \partial_i r)=\frac{r}{2}(1-N^2(r, t)).
\end{equation}
For convenience, in the context we use the units $G=\hbar=c=k_B=1$.

\section{non-equipartition thermodynamics relation in static case}
In this section, we recast the gravity into non-equipartition
thermodynamics relation in a static background. In \cite{Cao1}, a
Smarr-like thermodynamics relation has been considered on trapping
horizon \footnote{Although in \cite{Cao1}, this relation is
argued to be valid in the untrapped region, they also argue that the
Tolman temperature there is not propotional to the acceleration of
Kodama observers. Thus, we do not think this relation is
self-consistently valid in the untrapped region.} in spherical
symmetric case,
\begin{equation}
E=\frac{1}{2}A T +3 w V.\label{Cai}
\end{equation}
Here $E$ is the Misner-Sharp energy on trapping horizon, $A=4\pi r^2$ is the total number of `bits' on trapping horizon, $T=\kappa_h / 2\pi$ is the Hawking temperature corresponding to Kodama surface gravity, and $w=-T^a_a/2$ is the `work density'.

This relation can be extended to any virtual sphere of constant $r$ far from horizon. Consider a 3-hypersurface $\Sigma$ with a boundary $\partial \Sigma$ in a static spacetime with the metric components $\Phi=\Phi(r)$ and $N=N(r)$. The Kodama observer endures a 4-acceleration,
\begin{equation}
a^\mu=k^\nu \nabla_\nu k_\mu=(0,(N \partial_r N -N^2 \partial_r \Phi),0,0).
\end{equation}
According to the textbook \cite{Wald} , in a general space time, there is an useful differential geometric identity,
\begin{equation}
R_{\mu\nu}n^{\mu}n^{\nu}=\nabla_\rho (a^\rho-n^\rho\nabla_\lambda n^\lambda )-K_{\mu\nu}K^{\mu\nu}+K^2 . \label{id}
\end{equation}
Here $R_{\mu\nu}$ is the Ricci tensor, $K_{\mu\nu}$ is the extrinsic curvature of $\Sigma$, and $K$ is the trace of $K_{\mu\nu}$. In static case, Kodama vector parallels to Killing vector, and the extrinsic curvature terms vanish. So, the identity reduces to,
\begin{equation}
R_{\mu\nu}n^{\mu}n^{\nu}=\nabla_\rho a^\rho.
\end{equation}
The covariant 4-derivative in static spacetime are equivalent to,
\begin{equation}
R_{\mu\nu}n^{\mu}n^{\nu}=\nabla_\rho a^\rho=\frac{e^{\Phi}}{N}D_i(e^{-\Phi}N a^i).
\end{equation}
Here, $D$ is the covariant 3-derivative corresponding to the 3-space of the spacetime. Integrate it on both sides, and using Einstein equation and Gauss theorem, we will get the equipartition relation.
\begin{equation}
E_{Komar}=\frac{1}{4\pi} \int_\Sigma d^3 x \sqrt{h} R_{\mu\nu} \xi^\mu n^\nu=\int_\Sigma d^3 x \sqrt{h} \{D_i(e^{-\Phi}N a^i)\}=\frac{1}{2}\int_{\partial \Sigma}  \frac{d^2x \sqrt{\sigma}}{L^2_P}\Big\{\frac{e^{-\Phi} N a^i n_i}{2\pi}\Big\}.
\end{equation}
Here $\xi^\mu=e^{-\Phi}N n^\mu$ is the Killing vector, and $E_{Komar}$ is the Komar mass energy. $\sigma$ is the determinant of induced metric on $\partial \Sigma$ and $n_i$ is the unit spacial normal to $\partial \Sigma$. This relation can be viewed as the equipartition principle, if we define a Tolman temperature to Killing observer as $T_\xi=e^{-\Phi}NT_{loc}=e^{-\Phi}N a^i n_i/2\pi$. The number of degrees of freedom on $\partial \Sigma$ is $\Delta n=\sqrt{\sigma} d^2x /L^2_P$. So, it is recast as \cite{Pad3},
\begin{equation}
E_{Komar}=\frac{1}{2}\int_{\partial \Sigma}  dn T_\xi. \label{13}
\end{equation}
This thermodynamics relation is valid on any virtual sphere of constant $r$, but it is hard to be extended to the non-static case when Killing observer is absent. In order to get the thermodynamics relation in non-static case, first we need to replace Killing observer by Kodama observer in static case.
\begin{equation}
\begin{split}
\frac{1}{4\pi}\int_\Sigma d^3 x \sqrt{h}  R_{\mu\nu}K^{\mu}n^{\nu}
&=\frac{1}{4\pi}\int_\Sigma d^3 x \sqrt{h} \{D_i(N a^i)-\partial_r \Phi (\frac{\partial_r N}{N}-\partial_r \Phi) N^3 \} \\
&=\frac{1}{2}\int_{\partial \Sigma}  \frac{d^2x \sqrt{\sigma}}{L^2_P}\Big\{\frac{N a^i n_i}{2\pi}\Big\} -\frac{1}{4\pi}\int_\Sigma d^3 x \sqrt{h} \partial_r \Phi (\frac{\partial_r N}{N}-\partial_r \Phi) N^3.\label{17}
\end{split}
\end{equation}
In above equation we have used,
\begin{equation}
R_{\mu\nu}K^{\mu}n^{\nu}=D_i(N a^i)-\partial_r \Phi (\frac{\partial_r N}{N}-\partial_r \Phi) N^3.
\end{equation}
As a result , it is necessary to define a Kodama temperature as,
\begin{equation}
T_K =NT_{loc}=\frac{N a^i n_i}{2\pi}=\frac{N \partial_r N - N^2 \partial_r \Phi}{2\pi}.
\end{equation}
Notice that $T_K$ differs from $T_\xi$ by a redshift factor $e^{-\Phi}$. In Visser's definition of \cite{Abreu:2010ru, Abreu:2010sc}, the Tolman temperature $e^{-\Phi}T_K$ of Kodama observer coincides with $T_\xi$ in static case. However, in Hayward and Cai's definition \cite{Hayward1, Cao1}, the Hawking temperature of Kodama observer is mainly defined on trapping horizon,
\begin{equation}
T_K^\prime =\frac{1}{4\pi\sqrt{-\gamma}}\partial_i(\sqrt{-\gamma}\gamma^{ij}\partial_j r)|_h=\frac{N \partial_r N - \frac{1}{2}N^2 \partial_r \Phi}{2\pi}|_h.\label{T}
\end{equation}
It easy to see that $T_K=T_K^\prime$ on the black hole horizon due to $N^2=0$. In tunneling method \cite{CriscienzoPLB, CriscienzoCQG}, the tunneling rates observed by Kodama vector also produces the Hawking temperature $T_K^\prime$. Hence, in our article, we define Kodama temperature $T_K$ in Eq. (\ref{T}) as the Tolman temperature related to Kodama observer in thermodynamics. And $T_K^\prime$ can be regarded as a special case of $T_K$ on trapping horizon.

A recent work \cite{Wu:2010ty} uses a new holographic screen temperature by,
\begin{equation}
T_H =\frac{n^i \nabla_i e^{\psi}}{2\pi}=\frac{N\partial_r N}{2\pi},
\end{equation}
which is also defined in \cite{Cao1} on trapping horizon. Here $\psi$ is the generalized Newton potential defined by $e^{2\psi}=-K^\mu K_\mu$. It is also equivalent to $T_K$ and $T_K^\prime$ on trapping horizon. In our context, we don't have any holographic screens, and we only consider the correspondence between gravity and thermodynamics on virtual sphere of constant $r$. The relation between holographic screens and virtual sphere of constant $r$ will be discussed in Sec. 5.

The analogy of Komar mass energy in Eq. (\ref{17}) is not a well-defined conserved energy observed by Kodama observer. Thus, we need to replace it by Misner-Sharp energy,
\begin{equation}
\begin{split}
E_{MS}&=\int_\Sigma d^3 x \sqrt{h}  T_{\mu\nu}K^{\mu}n^{\nu}=\frac{1}{8 \pi}\int_\Sigma d^3 x \sqrt{h}  [R_{\mu\nu}-\frac{1}{2}g_{\mu\nu}R]K^{\mu}n^{\nu} \\
&=\frac{2}{8 \pi}\int_\Sigma d^3 x \sqrt{h}  R_{\mu\nu}K^{\mu}n^{\nu}-\frac{1}{8 \pi}\int_\Sigma d^3 x \sqrt{h}  [R_{\mu\nu}+\frac{1}{2}g_{\mu\nu}R]K^{\mu}n^{\nu}\\
&=\frac{1}{2}\int_{\partial \Sigma}  dn T_K  -\frac{1}{4 \pi}\int_\Sigma d^3 x \sqrt{h} \partial_r \Phi (\frac{\partial_r N}{N}-\partial_r \Phi) N^3-\frac{1}{8 \pi}\int_\Sigma d^3 x \sqrt{h}  [R_{\mu\nu}+\frac{1}{2}g_{\mu\nu}R]K^{\mu}n^{\nu}. \label{key_c0}
\end{split}
\end{equation}

This thermodynamics relation is our key conclusion in this section, which is an extension of the trapping horizon relation (\ref{Cai}), and an extension of the equipartition law \cite{Pad3}. In fact, it is also valid in spacetime without a horizon. When there is a black hole in the center of spacetime, this relation is also useful if we let $r_\Sigma \geqslant r_h$. When $\Phi=0$, it is equivalent with the equipartition rule Eq. (\ref{13}).

The physical interpretation of Eq.(\ref{key_c0}) is as follows. Defining the diagonal part of  $T^\mu_\nu$ as $-\rho(r)$, $P(r)$, $P_\theta(r)$, and $P_\phi(r)$ with $P_\theta(r)=P_\phi(r)$, the Misner-Sharp energy $E_{MS}$ can be expressed as,
\begin{equation}
E_{MS}=-\int_\Sigma d^3 x \sqrt{h}N T^0_0=4\pi\int_\Sigma \rho(r) r^2dr =\int_\Sigma \rho(r) dV.
\end{equation}
The analogy of Komar mass energy by Kodama vector is,
\begin{equation}
\begin{split}
\frac{1}{4 \pi}\int_\Sigma d^3 x \sqrt{h}  R_{\mu\nu}K^{\mu}n^{\nu}&=2\int_\Sigma d^3 x \sqrt{h}  (T_{\mu\nu}-\frac{1}{2}g_{\mu\nu}T)K^{\mu}n^{\nu} \\
&=\int_\Sigma (\rho(r)+P(r)+P_\theta(r)+P_\phi(r)) dV
\end{split}
\end{equation}
As a result, we can define a `pressure density' as,
\begin{equation}
w_2=\frac{1}{8\pi}\sqrt{g_{rr}}(R_{\mu\nu}+\frac{1}{2}g_{\mu\nu}R)K^{\mu}n^{\nu}=P(r)+P_\theta(r)+P_\phi(r).
\end{equation}
And then the last term of Eq.(\ref{key_c0}) is a volume integral of `pressure density' $w_2$.

The middle term of Eq.(\ref{key_c0}) is caused by the redshift factor $e^{\Phi}$, which is also the difference between Killing vector and Kodama vector. So, we define a `redshift work density' as,
\begin{equation}
w_1=\frac{1}{4\pi}\sqrt{g_{rr}}(\partial_r \Phi (\frac{\partial_r N}{N}-\partial_r \Phi) N^3)=\frac{\partial_r \Phi}{2}\frac{N\partial_r N-N^2\partial_r \Phi}{2\pi}=\frac{\partial_r (e^{\Phi})}{2}e^{-\Phi}T_K.
\end{equation}
When $\Phi=0$, this `redshift work density' term vanishes. So, we can re-express Eq.(\ref{key_c0}) in a simple form,
\begin{equation}
E_{MS}=\frac{1}{2}\int_{\partial \Sigma}  dn T_K  -\int_\Sigma (w_1+w_2) dV . \label{key_c}
\end{equation}
On trapping horizon, it is equivalent to Eq. (\ref{Cai}), where the `work density' term $3wV$ is composed of integrals of `redshift work density' $w_1$ and `pressure density' $w_2$. As a result, it is the extension of works in \cite{Pad3, Cao1} in static case.

\subsection{RN black hole}
For an RN black hole, we have $\Phi=0$ and $N^2=1-2M/r+Q^2/r^2$. Then we have the thermodynamics variables,
\begin{equation}
E_{MS}=\frac{r}{2}(1-N^2)=M-\frac{Q^2}{2r}.
\end{equation}
\begin{equation}
T_K=T_\xi=\frac{M-\frac{Q^2}{r}}{2\pi r^2}.
\end{equation}
\begin{equation}
w_1=0.
\end{equation}
\begin{equation}
w_2=\frac{1}{8\pi}\frac{Q^2}{r^4}.
\end{equation}
Therefore, the Eq.(\ref{key_c}) writes as,
\begin{equation}
E_{MS}=\frac{1}{2}(4\pi r^2)\frac{M-\frac{Q^2}{r}}{2\pi r^2}-\int_\Sigma dr \frac{1}{2}\frac{Q^2}{r^2}.
\end{equation}
Here $w_2$ is singular at $r=0$, so we need change the lower limit into $r=\epsilon$ ($\epsilon \rightarrow 0$) and add a integral constant $Q^2/2\epsilon$ to ensure the validity of the equation. This relation holds on any virtual sphere of constant r out of black hole horizon ($r\geqslant r_h$). When $r=r_h$, it reproduces Eq.(\ref{Cai}), and the integral of `pressure density' is equivalent to the `work density' term in Eq.(\ref{Cai}).

\subsection{Dilaton-Maxwell-Einstein black hole}
The charged stringy black hole is a non-vacuum solution of Einstein-Maxwell dilaton gravity in the string frame \cite{CriscienzoPLB, CriscienzoCQG}
\begin{equation}
ds^2=-(\frac{1-a/r}{1-b/r})dt^2+\frac{dr^2}{(1-a/r)(1-b/r)}+r^2d\Omega^2.
\end{equation}
In this case, we have $\Phi=\ln (1-b/r)$, $N^2=(1-(a+b)r+ab/r^2)$ and $a>b>0$. The horizon radius is $r=a$. The extremal limit as defined by a global structure is $b \rightarrow a$. However, the Killing temperature on trapping horizon is,
\begin{equation}
T_\xi=\frac{1}{4a\pi},
\end{equation}
which does not vanish in this limit. Since extremal black holes are expected to have zero-temperature, the Kodama temperature on trapping horizon,
\begin{equation}
T_K=\frac{a-b}{4a^2\pi},
\end{equation}
is a more general temperature. This puzzle are explained as the effect of gravitational dressing \cite{CriscienzoPLB, CriscienzoCQG}. Thus, it is reasonable to apply $T_K$ to investigate the thermodynamics,
\begin{equation}
E_{MS}=\frac{r}{2}(1-N^2)=\frac{1}{2}(a+b-ab/r).
\end{equation}
\begin{equation}
T_K=\frac{a-b}{4\pi r^2}.
\end{equation}
\begin{equation}
w_1=\frac{b}{2r(r-b)}\frac{a-b}{4\pi r^2}.
\end{equation}
\begin{equation}
w_2=\frac{1}{8\pi}\frac{b(ab/r-2a+b)}{(r-b)r^3}.
\end{equation}
\begin{equation}
w_1+w_2=-\frac{ab}{8\pi r^4}.
\end{equation}
Therefore, we have,
\begin{equation}
E_{MS}=\frac{1}{2}(4\pi r^2)\frac{a-b}{4\pi r^2}+\int_\Sigma dr \frac{1}{2}\frac{ab}{r^2}.
\end{equation}
Notice that the last integral can not vanish when $r \rightarrow \infty$, so it has an integral constant $b+ab/2\epsilon$.

It can be verified that the `work density' term $3wV$ in Eq.(\ref{Cai}), is equivalent to the `redshift work density' term plus the `press density' term on horizon in Eq.(\ref{key_c}).

\section{non-static case}
In non-static case, which has $\Phi=\Phi(t,r)$ and $N=N(t,r)$, the identity \ref{id} reduces to,
\begin{equation}
R_{\mu\nu}n^{\mu}n^{\nu}=\nabla_\rho (a^\rho-K n^\rho).
\end{equation}
Here $K$ is the trace of extrinsic curvature. As a result, the thermodynamics relation in non-static case is modified as,
\begin{equation}
E_{MS}=\frac{1}{2}\int_{\partial \Sigma}  dn T_K  -\int_\Sigma (w_1+w_2+w_3) dV . \label{key_c2}
\end{equation}
Here $w_3$ is the `gravitational work density' due to the non-vanishing extrinsic curvature,
\begin{equation}
w_3=\frac{1}{4\pi}\nabla_\rho (Kn^\rho) =-\frac{1}{4\pi}e^{2\Phi}\Big\{\frac{-3 (\partial_t N)^2}{N^4}+\frac{\partial_t^2 N}{N^3}+\frac{\partial_t \Phi \partial_t N}{N^3}\Big\}. \label{w_3}
\end{equation}
\subsection{Vaidya black hole}
Vaidya black hole is an approximate solution for an evaporating spherical symmetric black hole \cite{Vaidya},
\begin{equation}
ds^2=-(\frac{\dot m(t,r)}{f(m)})^2(1-\frac{2m(t,r)}{r})dt^2+\frac{dr^2}{1-\frac{2m(t,r)}{r}}+r^2d\Omega^2.
\end{equation}
Here, $m(t,r)$ is a slowly varying mass term, $\dot m=\partial_t m(t,r)$, and $f(m)$ is a term related to the Hawking evaporation. In this case, we have $\Phi=\ln \frac{f(m)}{\dot m}$, and $N^2=(1-2m/r)$. Then we have,
\begin{equation}
E_{MS}=\frac{r}{2}(1-N^2)=m.
\end{equation}
\begin{equation}
\partial_r \Phi=\Phi^\prime=\frac{f^\prime}{f}-\frac{\dot m^\prime}{\dot m}.
\end{equation}
\begin{equation}
T_K=\frac{m-rm^\prime-r(r-2m)\Phi^\prime}{2\pi r^2}.
\end{equation}
\begin{equation}
w_1=\frac{\Phi^\prime}{2}\frac{(m-rm^\prime-r(r-2m)\Phi^\prime)}{2\pi r^2}.
\end{equation}
\begin{equation}
w_3=\frac{e^{2\Phi}r}{4\pi (r-2m)^3}\{4\dot m^2+(r-2m)\dot m \dot \Phi+(r-2m) \ddot m \}.
\end{equation}
\begin{equation}
w_2=-w_3+ \frac{r^2(\Phi^{\prime2}-\Phi^{\prime\prime})+r(3m^\prime\Phi^\prime-m^{\prime\prime}-2m\Phi^{\prime2}+2m\Phi^{\prime\prime}-2\Phi^\prime)-m^\prime+m\Phi^\prime}{4\pi r^2}.
\end{equation}
\begin{equation}
w_1+w_2+w_3= \frac{-r^2\Phi^{\prime\prime}+r(2m^\prime\Phi^\prime-m^{\prime\prime}+2m\Phi^{\prime\prime}-2\Phi^\prime)-m^\prime+2m\Phi^\prime}{4\pi r^2}.
\end{equation}
Where, $\prime$ means partial derivative respect to $r$. Obviously, it satisfies,
\begin{equation}
E_{MS}=\frac{1}{2}(4\pi r^2)T_K-\int_\Sigma (w_1+w_2+w_3) dV.
\end{equation}
On the trapping horizon, it is equivalent to Eq. (\ref{Cai}). As a result, in the general non-static case, the `work density' term $3wV$ used by \cite{Hayward1, Cao1, Wu:2010ty}, are composed of the contributions from volume integrals on `redshift work density', `pressure density' and `gravitational work density'.

\subsection{FRW universe}
The FRW coordinate is an orthogonal co-moving coordinate, with $(t,r,\theta,\phi)$. Here `$t$' is the co-moving time, corresponding to a co-moving observer. The FRW metric is,
\begin{equation}
ds^2=-dt^2+\frac{a^2(t)}{1-kr^2}dr^2+a^2(t)r^2 d\Omega^2.
\end{equation}
$k$ is the intrinsic spacial curvature. The corresponding Kodama surface gravity is,
\begin{equation}
\kappa_H=\frac{1}{2\sqrt{-\gamma}}\partial_i(\sqrt{-\gamma}\gamma^{ij}\partial_j )(a(t)r)|_h=-\frac{1}{2}(2H^2+\dot H+\frac{k}{a^2})(a(t)r)|_h.
\end{equation}
Thus, we hope that the thermodynamics relation still hold for FRW universe. In order to apply Eq.(\ref{key_c2}), we need to make a general coordinate transformation to Kodama observer coordinate. Setting $R=a(t)r$, and $T=T(t,r)$ which satisfy,
\begin{equation}
\begin{split}
dT&=\frac{1-\frac{k}{a^2}R^2}{1-H^2R^2-\frac{k}{a^2}R^2}dt+\frac{aHR}{1-H^2R^2-\frac{k}{a^2}R^2}dr,\\
dR&=HRdt+adr. \label{trans}
\end{split}
\end{equation}
Then, we will get a new orthogonal coordinate,
\begin{equation}
ds^2=-\frac{1-H^2R^2-\frac{k}{a^2}R^2}{1-\frac{k}{a^2}R^2}dT^2 +\frac{dR^2}{1-H^2R^2-\frac{k}{a^2}R^2} +R^2 d\Omega^2.
\end{equation}
Note that the covariant divergence $\kappa_H$ is not invariant under coordinate transformation (\ref{trans}), so the transformation is not physical unless
\begin{equation}
\partial_t \partial_r T(t,r)=\partial_r \partial_t T(t,r),
\end{equation}
for any $t, r$. Then the constraint equation is,
\begin{equation}
\partial_r (\frac{1-kr^2}{1-\dot a^2r^2-kr^2})=\partial_t (\frac{a\dot a r}{1-\dot a^2r^2-kr^2}).
\end{equation}
Solving this equation, we have,
\begin{equation}
r(a\ddot a-\dot a^2)-r^3(\dot a^2(\dot a^2-a\ddot a)+k(a\ddot a+\dot a^2))=0.
\end{equation}
The coefficients of $r$ and $r^3$ must vanish at any $(t, r)$. So, the only physical transformation must satisfy,
\begin{equation}
\begin{split}
a&=e^{Ht}+c, \\
k&=0,
\end{split}
\end{equation}
where $H, c$ are constant and $H=\dot a/a$ is the Hubble parameter of FRW universe. Therefore, there is no physical relation between FRW co-moving observer and Kodama observer, unless it is a de-Sitter universe. In this case, Eq.(\ref{key_c2}) are trivially satisfied in a vacuum solution. In the other cases of FRW universe, since the Kodama preferred time flow dose not exist, it is so far suspicious to define thermodynamics variables directly from Kodama observers. And we should be more careful when we investigate relation of thermodynamics and gravity in the real universe.

\section{The differential form}
The differential form of thermodynamics directly comes from the radical-temporal part of Einstein equation \cite{Hayward1, Cao1, Chen:2010ay, Tian, Wu:2010ty}, in the general case which is,
\begin{equation}
G_0^0 =\frac{1}{r^2}(rf^\prime-(1-f))=8\pi T_0^0,
\end{equation}
\begin{equation}
G_0^1 =-\frac{\dot f}{r}=8\pi T_0^1,
\end{equation}
\begin{equation}
G_1^1 =\frac{1}{r^2}(rf^\prime-2rf\Phi^\prime-(1-f))=8\pi T_1^1,\label{radical}
\end{equation}
where we denote $f(t,r)=N^2(t,r)$ for simplicity. When $\Phi=0$ and $f=f(r)$, the above equations reduce to,
\begin{equation}
G_0^0=G_1^1 =\frac{1}{r^2}(rf^\prime-(1-f))=-8\pi\rho=8\pi P.
\end{equation}
Multiplying the radical part by $r^2dr/2$, we will find the first law of holographic screen thermodynamics,
\begin{equation}
TdS-dE=PdV.
\end{equation}
Here $T=T_K=f^\prime/4\pi$ is the holographic screen temperature, $S=\pi r^2$ is the holographic screen entropy, $E=r(1-f)/2$ is the Misner-Sharp mass at $f=$const, and $V$ is the coordinate volume. This first law describes the thermodynamics of different states of one holographic screen.

When $\Phi=\Phi(r)$ and $f=f(r)$, the Einstein equations reduce to,
\begin{equation}
G_0^0+G_1^1 =\frac{1}{r^2}(2rf^\prime-2rf\Phi^\prime-2(1-f))=8\pi(P-\rho), \label{diff}
\end{equation}
Multiplying the radical part by $r^2dr/4$ and evaluate it on $f=0$, we will find the first law of thermodynamics \cite{Hayward1, Cao1},
\begin{equation}
T_h dS_h-dE=-wdV.
\end{equation}
Where $T_h=T_k^\prime=\frac{f^\prime-f\Phi^\prime}{4\pi}$ is the Hawking temperature on trapping horizon define by Hayward, and $w=(\rho-P)/2$ is the `work density'.

Multiplying the radical part by $r^3/4$, we will find the Smarr-like law \cite{Cao1},
\begin{equation}
2T_h S_h-E=-3wV.
\end{equation}

Analogous to this, in the general case $\Phi=\Phi(t,r)$ and $f=f(t,r)$, Eq. (\ref{radical}) can be written as,
\begin{equation}
T_K dS-dE=PdV, \label{1law}
\end{equation}
and
\begin{equation}
2T_K S-E=3PV,
\end{equation}
which seems to be an extension of first law and Smarr-like law in the general case. However, there are several limitations of the extension. First, the equations is lack of the time-derivatives terms, which is the `grativational work density' in the integral form (\ref{key_c2}). So, this differential first law does not describe all the thermodynamics.

Second, Eq. (\ref{1law}) is only valid on the definition of holographic screen $f=$const. While $T_K$ is Tolman temperature defined on virtual sphere $r=$const. In the static case, $f=$const is equivalent to $r=$const, so the Kodama observers are at rest on the holographic screens. In this case, Eq. (\ref{1law}) can be well interpreted consistently. However, in the non-static case, when $f=$const, $r$ varies. It means the Kodama observers are crossing the holographic screens eternally. In this case, how to interpret Eq. (\ref{1law}) in physics is still an open question, since $T_K$ and $E$ are not defined on the same virtual sphere.

The spherical part of Einstein equation is,
\begin{equation}
G_2^2=G_3^3 =e^{2\Phi}\frac{f\dot f \dot \Phi-2\dot f^2+f\ddot f}{2f^3}+\frac{f^\prime(2-3r\Phi^\prime)+rf^{\prime\prime}+2f(r\Phi^{\prime2}-\Phi^\prime-r\Phi^{\prime\prime})}{2r}=8\pi P_\theta=8\pi P_\phi.
\end{equation}
A general differential first law might be obtained the sum of all the non-vanishing diagonal part of Einstein equation, which does not give more information than Eq. (\ref{key_c2}). Furthermore, in order to well define the thermodynamical variables on holographic screens, one might introduce a new congruence of observers co-moving with holographic screens other than Kodama observer. It will be far more complicated than Eq. (\ref{key_c2}). Notice that Eq. (\ref{key_c2}) are evaluated on the hypersurface $t=$const, and the $r=$const virtual sphere out of trapping horizon can be regarded as holographic screens in the same moment. As a result, Eq. (\ref{key_c2}) is so far a well-defined, simple and complete description of holographic screen thermodynamics from the gravity. And we are not seeking any differential first law in this article.

\section{Conclusion}
Although previously, there are some discussions on thermodynamics by Kodama observers in general non-static case, these thermodynamics can not be extended from trapping horizon to untrapped region, for the previous Tolman temperature is not defined consistently in that region. In this work, by defining a new Tolman temperature of Kodama observer, we give a generalized thermodynamics relation (\ref{key_c2}) on virtual sphere of constant $r$ in non-static spherical symmetric spacetimes. This relation contains work terms contributed by `redshift work density', `pressure density' and `gravitational work density', which are equivalent to the `work density' in \cite{Cao1}. As a result, it can be regarded as extension of \cite{Pad3, Cao1}. We illustrate it in RN black hole, Dilaton-Maxwell-Einstein black hole and Vaidya black hole. We also find that it can not describe the FRW universe unless in the vacuum case. Seeking for a generalized differential form of first law is difficult due to the obscure definition of holographic screen variables, and it would not give more information than (\ref{key_c2}).

The formulism of (\ref{key_c2}) can be easily extended into higher-dimensional gravity theories like Gauss-Bonnet, Lovelock, or $f(R)$ gravity. Reversing the logic of this article, one may find a generalization of entropic force in non-static spacetimes. We will investigate the ideas in future.

\acknowledgments

We thank K.N.Shao, W.J.Jiang, Z. Yin, Q.J.Cao and Y. Wang for useful discussions. Chen and Li are supported in
part by the NNSF of China Grant No. 10775116, 973 Program Grant No.
2005CB724508, and ``the Fundamental Research Funds for the Central
Universities". Wang is supported by the Fundamental Research Funds for the Central Universities (No. lzujbky-2009-122)
and the Fundamental Research Fund for Physics and Mathematics of Lanzhou University (No. LZULL200912). Chen would like to thank the organizer and the
participants of the advanced workshop, ``Dark Energy and Fundamental
Theory" supported by the Special Fund for Theoretical Physics from
the National Natural Science Foundation of China with grant no:
10947203, for stimulating discussions and comments.

\end{document}